\documentstyle[epsfig,here,12pt]{article}

\newcommand {\eqref} [1] {(\ref {#1})}
\newcommand {\beq} {\begin{equation}} 
\newcommand {\eeq} {\end{equation}}
 \newcommand {\ber}{\begin{eqnarray*}}
 \newcommand {\eer} {\end{eqnarray*}}
\newcommand {\bea}{\begin{eqnarray}}
 \newcommand {\eea} {\end{eqnarray}}

\def\Acknowledgements{\bigskip  \bigskip {\begin{center} \begin{large}
             \bf ACKNOWLEDGEMENTS \end{large}\end{center}}}

\begin{document}\begin{titlepage}
\rightline{{CPTH-S055.0500}}
\vskip 1cm
\centerline{{\Large \bf Comments on Perturbative Dynamics of}}
\vskip 0.2cm
\centerline{{\Large \bf Non-Commutative Yang-Mills Theory}}
\vskip 1cm
\centerline{Adi Armoni}
\vskip 0.5cm
\centerline{Centre de Physique Th{\'e}orique de l'{\'E}cole 
Polytechnique}
\centerline{91128 Palaiseau Cedex, France}
\vskip 0.5cm
\centerline{armoni@cpht.polytechnique.fr}
\begin{abstract}
We study the $U(N)$ non-commutative Yang-Mills theory at the one-loop
approximation. We check renormalizability and gauge invariance of the
model and calculate the one-loop beta function.
The interaction of the $SU(N)$ gauge bosons with the $U(1)$ gauge
boson plays an important role in the consistency check. In
particular, the $SU(N)$ theory by
itself is not consistent. We also find that the $\theta \rightarrow 0$
 limit of the $U(N)$ theory does not converge to the ordinary
 $SU(N)\times U(1)$ commutative theory, even at the planar limit.
 Finally, we comment on the UV/IR mixing.
\end{abstract}
\end{titlepage}

\section{Introduction and Conclusions}

Non-commutative gauge field theories lately attracted a lot of attention,
 mainly due to the discoveries of their relation to string theory \cite{SW}.
It was also found that the perturbative structure of these theories 
has an interesting pattern. It was shown \cite{Filk}, in the case of
 scalar theory, that planar diagrams
 of the non-commutative theory are the same as planar diagrams of
 ordinary commutative theory, up to global phases. For an earlier
 related work see ref.\cite{GO}. It was then suggested
 \cite{BS} that non-planar graphs are UV finite, due to the oscillatory
Moyal phase which regulates the integrals. It was
 found later \cite{MRS} that these contributions actually lead to
  divergences, which were interpreted as {\em infra-red} divergences.
These contributions are singular in the $\theta \rightarrow 0$ limit
 and they occur also in gauge theories \cite{Hayakawa,MST}.

This paper is devoted to the study of $U(N)$ non-commutative gauge
theory. The $U(1)$ case was already studied by several authors
\cite{MaSa,KW,Jabbari,Hayakawa,MST}.
 The renormalization of the model, at the one loop 
approximation, was studied first in \cite{MaSa}. The UV/IR mixing, in the
$U(1)$ case was studied by \cite{Hayakawa} and \cite{MST}. Related
works about perturbative dynamics of non-commutative field theories are
\cite{CR,Aref1,GKW,Aref2,RS,AS,GM2,CHU,Aref3,RR,FI,BGPS,GM}. Perturbative
aspects of non-commutative field theories from string theory were
discussed in \cite{AD,KL,BCR,GKMRS,LM,CRS}.

The non-commutative $U(N)$ Yang-Mills action is
\beq
\int d^4 x\ {\rm tr} \  -{1\over 2g^2} F_{\mu \nu} \star F^{\mu \nu} \label{YM}
\eeq
where $F_{\mu \nu}$ is
\beq
F_{\mu \nu} = \partial _\mu A_\nu - \partial _\nu A_\mu -i( A_\mu
\star A_\nu - A_\nu \star A_\mu)
\eeq
and $A_\mu$ is a $N\times N$ matrix. The $\star$-product between
two functions $f$ and $g$ is defined by
\beq
f\star g (x) = e^{{i\over 2} \theta ^{\mu \nu} \partial ^{(\xi)} _\mu \partial
^{(\eta)} _\nu} f(x+\xi) g(x+\eta) |_{\xi,\eta \rightarrow 0}.
\eeq 
The action \eqref{YM} is invariant under $U(N)$ gauge transformation
\beq
\delta _\lambda A_\mu = \partial _\mu \lambda - i(A_\mu \star \lambda -
\lambda \star A _\mu).\label{GT} 
\eeq
The gauge transformation \eqref{GT} is different from the commutative
 gauge transformation in the sense that it mixes the $U(1)$ gauge
boson with the $SU(N)$ gauge bosons. In fact, the non-commutative 
Yang-Mills action \eqref{YM} also mixes the $U(1)$ and the $SU(N)'s$. It cannot
 be written as a sum of a $SU(N)$ and a $U(1)$ theories as
 the ordinary YM theory, since there
are interaction terms between the $SU(N)$ gluons and the $U(1)$
'photon'. In order to demonstrate this point we list in figure 1 below
the Feynman rules which describe the 3-gluons interaction (the full
list of Feynman rules is written in Appendix A)

\begin{figure}[H]
  \begin{center}
\mbox{\kern-0.5cm
\epsfig{file=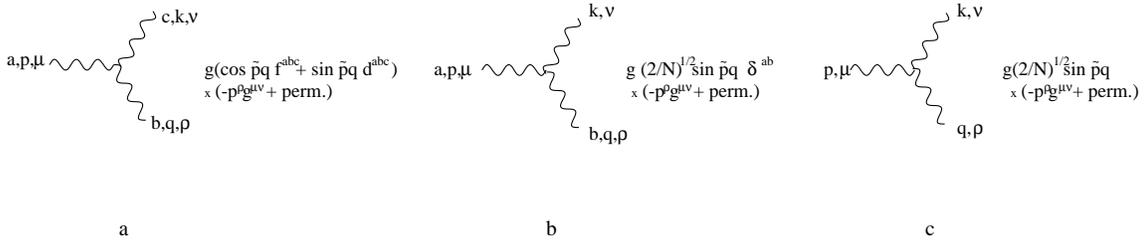,width=15.0true cm,angle=0}}
\label{rules2}
  \end{center}
\caption{Contributions to the 3 gluons vertex. a).
$SU(N)-SU(N)-SU(N)$ interaction. b). $SU(N)-SU(N)-U(1)$ interaction.
c). $U(1)-U(1)-U(1)$ interaction.}
\end{figure}

The complicated structure of the action \eqref{YM} raises the question
of gauge invariance and consistency of the non-commutative Yang-Mills
theory at the quantum level. The action
\eqref{YM} consists of many interaction terms with a single coupling
$g$ - due to gauge invariance. It is not clear, a-priori, that the
relations among the various couplings in the action is kept at the quantum
level. There are two limits of the theory which hint that the full
$U(N)$ gauge invariance might be broken. The first limit is
 $\theta \rightarrow 0$. In this case, the theory is expected to
reduce to the ordinary commutative theory. However, the commutative
theory has a $SU(N)\times U(1)$ gauge symmetry and the $SU(N)$
coupling is not related to the $U(1)$ coupling by gauge symmetry.
Moreover, at the quantum level the $SU(N)$ coupling runs and the
$U(1)$ coupling is kept fixed. The second limit, is the planar limit.
Since it looks as if the non-commutative theory and the commutative
 are identical at the planar level, the same question about the $U(1)$
coupling should be raised.

As we shall see the $U(N)$ gauge symmetry is not broken quantum
mechanically. The renormalization procedure does not violate the
relations between the various couplings (at least at the one loop
level). The resolution of the puzzles mentioned above, is the
following: the limit $\theta \rightarrow 0$ does not lead to the
ordinary commutative theory. Though the resulting action looks like 
the ordinary YM action (note that the $U(1)$ and the $SU(N)$ seems to 
decouple in the $\theta \rightarrow 0$ limit, see figure 1b), the $U(1)$ and 
$SU(N)$ couplings have exactly the same beta function.

The fact that the limit $\theta \rightarrow 0$ of the $U(1)$ theory is
singular was already pointed out in \cite{MST}. It was with
relation to the non-planar contributions, which are
manifestly singular in $\theta$. We claim, however, that the theory is
not smooth in $\theta$ even in the planar limit.
In this case, indeed the $SU(N)$ sector theory looks like the
commutative theory, but the interaction of the $U(1)$ with the $U(N)'s$
 survives the limit. In particular, the $U(1)$ gauge coupling runs. 
Thus, the planar sector of the $\theta \rightarrow
0$ theory does not correspond to the planar sector of the commutative theory. 
In this way, the puzzle about $U(N)$ gauge invariance at the quantum
level is also resolved.  

The main results of the paper are the following: in section 2 we
calculate the counter terms which are needed to regulate the
divergences in the planar graphs of the $SU(N)$ and $U(1)$ gluons
propagators. We find that they are the same and equal to the ordinary
commutative counter term of the $SU(N)$ propagator. The non-planar
contributions, however, are different. There is a non-planar finite
contribution to the $U(1)$ propagator\cite{MST},
 but there is no such contribution for the $SU(N)$.
 In section 3 we calculate the
 counter terms of the various 3 gluons vertices. Our results in this
section are similar to those of section 2. The divergent (planar) part of the
various 3 gluons vertices is the same, but the finite (non-planar) part is
different. Finally, in section 4, we calculate the beta function,
discuss our results and more general cases where also matter fields are
present. 

We shall use the following conventions: capital letters $(A,B,C,...)$
denote $U(N)$ indices, small letters $(a,b,c,...)$ denote $SU(N)$
indices. The $U(1)$ generator is normalized as follows
 $t^0={1\over \sqrt{2N}}$, such that
${\rm tr} \ t^A t^B = {1\over 2} \delta ^{AB}$. Finally,
$[t^a,t^b]=if^{abc}t^c$ and $\{ t^a,t^b \} = {1\over N} \delta ^{ab} +
d^{abc} t^c$. Thus $d^{abc}$ represents the symmetric tensor for the
 fundamental representation. In addition, we shall use the notation $\tilde
p _\mu = {1\over 2} \theta _{\mu \nu} p^\nu$.

\section{Corrections to the gluon propagator}
In order to check the renormalizability and gauge invariance at the
 quantum level, let us start with the one loop correction to the
 gluon propagator. The various contribution are drawn in figure 2 below

\begin{figure}[H]
  \begin{center}
\mbox{\kern-0.5cm
\epsfig{file=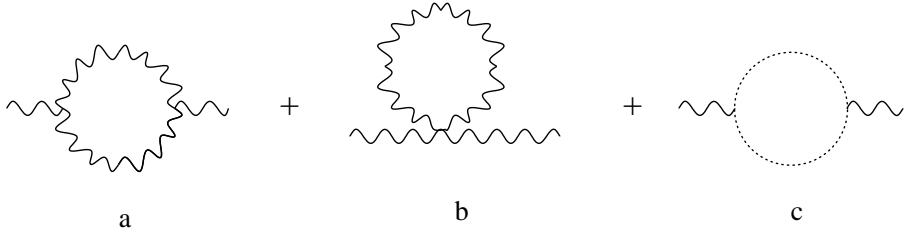,width=12.0true cm,angle=0}}
\label{prop}
  \end{center}
\caption{One loop corrections to the gluon propagator.}
\end{figure}

We consider first the case where the external legs carry $U(1)$
indices ('photons'). The calculation is a straightforward
generalization of the calculations which were performed for the $U(1)$
non-commutative Yang-Mills theory\cite{Hayakawa,MST}.
 The three contributes are drawn in
figure 2. Let us focus on diagram 2a. The only
 difference in comparison with the $U(1)$ theory is that now all
the $U(N)$ gluons can circulate in the loop.
 We will use the Feynman rules 1b and 1c.
We denote the external momentum by $p$ and the internal momentum by
$q$. The resulting expression is

\bea
A^{\mu \nu}
& = & {1\over 2} \int {d^4q\over (2\pi)^4} {-i\over q^2} {-i\over (p+q)^2}
\times T_1  \times
\nonumber 
\\
& &
 (g^{\mu \rho} (p-q)^\sigma + g^{\rho \sigma} (2q+p)^\mu + g^{\sigma
\mu} (-q-2p)^ \rho ) \times \nonumber \\
& &
(\delta ^\nu _\rho (q-p)_\sigma + g_{\rho \sigma}
 (-2q-p)^\nu + \delta ^\nu _\sigma (q+2p)_\rho ), \label{A11} \\
 T_1 & = & {2g^2 \over N} \delta ^{AB} \delta ^{AB} \sin ^2 \tilde pq
\eea 

By using the identity
\beq 
\sin ^2 \tilde pq = {1\over 2} (1-\cos 2\tilde pq) \label{trigo}
\eeq
we can isolate the planar contribution which comes from the ${1\over
2}$, from the non-planar contribution (the cosine)
\beq
 T_1 = g^2 N + \ \mbox{non-planar term.}
\eeq
 The planar part
of the contribution is divergent and its value is exactly the
 same as the value of the divergent part of the gluon propagator
 in ordinary $SU(N)$ Yang-Mills theory. The same pattern occurs
in the other diagrams in figure 2. Indeed, after the summation of the
three diagrams in figure 2 we find that in order to cancel the
divergent part of the $U(1)$ propagator the following counter term is
 needed
\beq
\delta ^{(1-1)} _3 = {g^2 N \over (4\pi)^2}\times {5\over 3} \times
{2\over \epsilon}, \label{CT11} 
\eeq 
where dimensional regularization was used and $\epsilon = 4-d$.
Note that the counter term does not depend on $\theta$. As long as
$\theta$ is non zero, a counter term \eqref{CT11} is needed. Otherwise
 the $U(1)$ theory is free. Therefore, though the Feynman rules of the
 theory are smooth in $\theta$, the limit $\theta \rightarrow 0$ is
singular.

For completeness let us quote the result for the finite part of the
correction to the $U(1)$
propagator \cite{MST}. It is calculated by replacing the cosine of \eqref{trigo} 
by an exponent and by looking at the high momentum regime in the
integrals of \eqref{A11} and the two other diagrams in figure 2. The result is \cite{MST} 
\beq
 A ^{\mu \nu} _{finite} =  -2g^2 N \int {d^4q \over (2\pi)^4} {2q^\mu
q^\nu - g^{\mu \nu} q ^2 \over
q^4 } e^{i2\tilde p q} \sim  
 g^2 N {\tilde p ^\mu \tilde p ^\nu \over \tilde p ^4}.
\eeq 

Note that this term is singular in $\theta$. When inserted into higher loops it behaves
as ordinary {\em 
infra-red} divergences \cite{MRS}. Thus, an effect which was
originally
 due to high momentum turns out to be an IR effect. This is the UV/IR mixing
which was found in \cite{MRS}.

Let us turn now to the calculation of the correction to the $SU(N)$
part of the gluon. Again, let us start with diagram 2a.
The coupling of the $SU(N)$ bosons which circulate in the loop
contains the symmetric tensor $d^{abc}$. The integral is the same
as \eqref{A11}, but $T_1$ is replaced by $T_2$ 
\beq
 T_2 =
g^2 (f^{xya} \cos \tilde pq + d^{xya} \sin \tilde pq)
    (f^{xyb} \cos \tilde pq + d^{xyb} \sin \tilde pq) 
\eeq 
By using the $SU(N)$ identities (see Appendix B)
\bea
& & f^{axy} f^{bxy} = N \delta ^{ab}  \label{I1} \\
& & d^{axy} d^{bxy} = (N-{4\over N}) \delta ^{ab} \label{I2}
\eea
$T_2$ can be written as 
\beq
T_2 =
g^2 N( 1 -{4\over N^2}\sin ^2 \tilde pq). \label{T2}
\eeq
 Interestingly there is another contribution to
the gluon propagator which doesn't occur in ordinary Yang-Mills
theory, due to the existence of new vertices (fig. 1b).
 It is possible to exchange a $U(1)$ boson in half
of the loop and $SU(N)$ boson in the other half. It contributes
\beq
 T'_2 = 2\times g^2 {2 \over N} \sin ^2 \tilde pq \label{T22}
\eeq
(the factor 2 in \eqref{T22} represents two possible exchanges of the
$U(1)$).
Collecting the two terms \eqref{T2} and \eqref{T22}, we find that the one loop correction
 to the $SU(N)$ propagator is exactly the same as in the commutative
 case. Thus the divergences can be compensated by the commutative
counter term
\beq
\delta ^{(N-N)} _3 = {g^2 N \over (4\pi)^2}\times {5\over 3} \times
{2\over \epsilon}. \label{CTNN} 
\eeq 
Remarkably \eqref{CT11} is identical to \eqref{CTNN} (except that
\eqref{CTNN} in needed also when $\theta =0$, in contrast to
\eqref{CT11}). This fact is
 crucial to ensure gauge invariance of the model at the quantum level, as we shall see later.

It is interesting to note that the corrections to the $SU(N)$
 propagator
 does not contain a non-planar finite part.
 Therefore, though the propagators are
identical in their divergent part, they differ in their
finite part. It is due to non-planar graphs which exist in the
corrections to the $U(1)$ propagator but do not exist for the $SU(N)$ one.   

\section{Corrections to the 3-gluons vertex}
In this section we calculate the one-loop corrections to the 3 gluons
vertex. The relevant diagrams are listed in figure 3 below. The
external momenta are $p_1,p_2,p_3$ with $p_1+p_2+p_3=0$. 
\begin{figure}[H]
  \begin{center}
\mbox{\kern-0.5cm
\epsfig{file=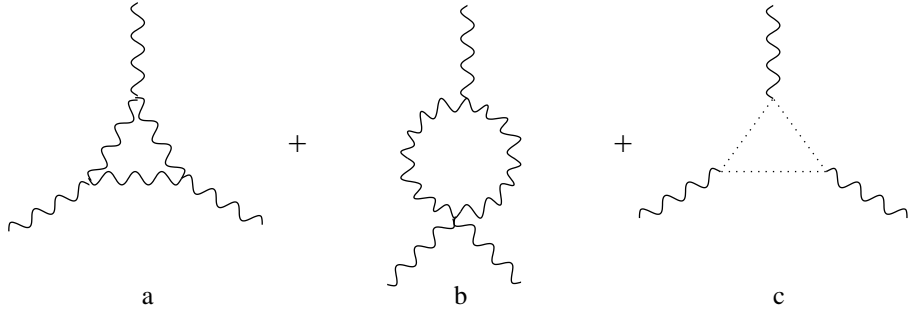,width=12.0true cm,angle=0}}
\label{vertex}
  \end{center}
\caption{One loop corrections to the 3-gluons vertex.}
\end{figure}

We begin with the simplest case in which all external legs are
$U(1)'s$. We focus on diagram 3a. The gluons which circulate in the
 loop belongs to $U(N)$. Similar calculations were made in
refs.\cite{Hayakawa,MST}.

\bea
 M^{\mu \nu \rho} & = & {1\over 2} \int {d^4q\over (2\pi)^4} {-i\over
q^2} {-i\over (q-p_2)^2} {-i\over (q+p_1)^2} \times V_1 \times \\
\nonumber & &
 (g^{\sigma _1 \mu} (q-p_1)^{\sigma _2} + g^{\mu \sigma _2} (2p_1+q)
^{\sigma _1} + g^{\sigma_2 \sigma_1} (-p_1-2q)^\mu ) \times \\
\nonumber & & 
(\delta ^\nu _{\sigma _1}  (p_2 + q )^{\sigma _3 } +
 \delta _{\sigma_1} ^{\sigma _3}
(-2q +p_2) ^\nu + g^{\sigma_3 \nu} (q-2p_2) _{\sigma _1} ) \times \\
& & 
(g_{\sigma _3 \sigma _2 } (-2q-p_1+p_2 )^{\rho } + \delta _{\sigma_2} ^\rho
(p_1+q-p_3) _{\sigma_3}  + \delta ^\rho _{\sigma_3} (p_3 + p_2 -q) _{\sigma_2}
), \nonumber \\
 V_1 & = & g^3 {({2\over N})} ^{3\over 2} \delta ^{XY} \delta ^{YZ} 
\delta ^{ZX} \sin \tilde p_1 q \sin \tilde p_2 q \sin \tilde p_3 (q+p_1)   
\eea

By using the following identity
\bea
& & \sin \tilde p_1 q \sin \tilde p_2 q \sin \tilde p_3 (q+p_1) =
\nonumber \\ 
& & -{1\over 4} \cos \tilde p_3 p_1 (\sin 2\tilde p_1 q + \sin 2\tilde
p_2 q + \sin 2\tilde
p_3 q) \nonumber \\
& & -{1\over 4} \sin \tilde p_3 p_1 (1 - \cos 2\tilde p_1 q - \cos
2\tilde p_2 q + \cos 2\tilde p_3 q) \label{sin}
\eea
we can isolate the divergent parts from the finite parts. The divergent
part comes from the $-{1\over 4} \sin \tilde p_3 p_1$ contribution.
Since this part does not depend on $q$ it would lead to a contribution which is
similar to the commutative case. The other diagrams in figure 3 are
similar. Thus the needed counter term to cancel the divergent part
 of the $U(1)-U(1)-U(1)$ vertex is
\beq
\delta _1 ^{(1-1-1)} = {2g^2 N \over (4\pi)^2} \times {1\over 4}
\times {4\over 3} \times {2\over \epsilon} = {g^2 N \over (4\pi)^2} \times
{2\over 3} \times {2\over \epsilon} \label{CT111}
\eeq
The finite part of the correction, which arise from the $-{1\over 4}
\cos \tilde p_3 p_1 \times \sin 2\tilde p_i q$ terms in \eqref{sin}, 
 is calculated by replacing the sin by an exponent. The procedure is
exactly the same as in the $U(1)$ case \cite{MST}. The result is
\bea
\lefteqn{M ^{\mu \nu \rho} _{finite} = } \nonumber \\
& & g^3 \sqrt{2N} \cos \tilde p_3 p_1 \int {d^4q \over (2\pi)^4} {1\over
q^6} (4q^\mu q^\nu q^\rho -q^2(q^\mu g^{\nu \rho} + q^\nu g^{\mu \rho}
+ q^\rho g^{\mu \nu} ))\times \nonumber \\
& & (e^{i 2\tilde p_1 q} + e^{i2\tilde p_2 q} +
e^{i 2\tilde p_3 q}) \nonumber \\
& &
\sim g^3 \sqrt{N} \cos \tilde p_3 p_1
\left ( 
{\tilde p_1 ^\mu \tilde p_1 ^\nu \tilde p_1 ^\rho \over \tilde p_1 ^4}+
{\tilde p_2 ^\mu \tilde p_2 ^\nu \tilde p_2 ^\rho \over \tilde p_2 ^4}+
{\tilde p_3 ^\mu \tilde p_3 ^\nu \tilde p_3 ^\rho \over \tilde p_3 ^4}
\right ) \label{finiteV} 
\eea
The calculation of the correction to the 3-gluons vertex when the
external legs are in $SU(N)$, is a bit more complicated, as there are
many contributions in the non-commutative case. The first contribution 
is when $SU(N)$ gluons circulate in the triangle of figure 3a. We will
use the Feynman rules in figure 1a. The calculation of the diagram is
performed by replacing $V_1$ by
\bea
 V_2 & = &
 g^3 (f^{axy} \cos \tilde p_1 q + d^{axy} \sin \tilde p_1 q) \times
\nonumber \\
& &     (f^{byz} \cos \tilde p_2 q + d^{byz} \sin \tilde p_2 q) \times
\nonumber \\
& &     (f^{czx} \cos \tilde p_3 (q+p_1) + d^{czx} \sin \tilde
p_3(q+p_1)) \label{V2}
\eea
In order to simplify \eqref{V2} we use the following $SU(N)$
identities (see Appendix B for derivation)
\bea
 & & f^{axy} f^{byz} f^{czx} - f^{axy} d^{byz} d^{czx} \nonumber \\
 & & -  d^{axy} f^{byz} d^{czx}  -  d^{axy} d^{byz} f^{czx} =
2N(1-{3\over N^2}) f^{abc} ,\\
& & d^{axy} d^{byz} d^{czx}  - d^{axy} f^{byz} f^{czx} \nonumber \\
& & -f^{axy} d^{byz} f^{czx} - f^{axy} f^{byz} d^{czx} =
2N(1-{3\over N^2}) d^{abc} ,
\eea
and trigonometric identities similar to \eqref{trigo}. Hence, $V_2$ can
 be written as follows
\beq
 V_2 = - g^3 (f^{abc} \cos \tilde p_3 p_1 \  + d^{abc} \sin \tilde p_3 p_1)
{N\over 2} (1 - {3\over N^2}) + \ \mbox{other terms} ,
\eeq
where 'other terms' means additional contributions which do not
 lead to divergences. 
Apart from the $V_2$ contribution, there is another contribution to
the $SU(N)-SU(N)-SU(N)$ vertex. It is due to $SU(N)$ bosons flowing in two
of the sides of the triangle in figure 3a and a $U(1)$ boson in the
third side.
The contribution is 
\beq 
V'_2 = 3\times g^3 {2\over N} 
 (f^{axy} \cos \tilde p_1 q + d^{axy} \sin \tilde p_1 q) \times
 \delta ^{xb} \sin \tilde p_2 q \times \delta ^{yc} \sin \tilde p_3(q+p_1)
\label{V22}
 \eeq
The part that leads to divergences in \eqref{V22} can be written as follows
\beq
 V'_2 = - g^3 (f^{abc} \cos \tilde p_3 p_1 \  + d^{abc} \sin \tilde p_3 p_1)
{N\over 2}  {3\over N^2}  
\eeq
Thus, collecting the two contributions $V_2$ and $V'_2$, we find that
 the counter term which is needed to cancel the divergences in the 3
gluons vertex with external legs in $SU(N)$ is
\beq
\delta _1 ^{(N-N-N)} = {g^2 N \over (4\pi)^2} 
\times {2\over 3} \times {2\over \epsilon} ,  \label{CTNNN}
\eeq
as in ordinary commutative Yang-Mills theory. Note that the
interaction with the $U(1)'s$ was needed to cancel the ${1\over N^2}$
terms in \eqref{V2}. Another comment is that the finite contribution
\eqref{finiteV} in the $U(1)-U(1)-U(1)$ cancels in the present case.

We turn now to the renormalization of 3-gluons vertex with one external
leg in $U(1)$ and two external legs in $SU(N)$ (figure 1b).
The first contribution to the diagram 3a is when $SU(N)$ bosons circulate
in the loop. We should use the Feynman rules 1a and 1b.
The contribution is 
\bea 
& & V_3 =  g^3 \sqrt{2\over N}  (f^{axy} \cos \tilde p_1 q +
 d^{axy} \sin \tilde p_1 q) \times \nonumber \\
& &     (f^{byz} \cos \tilde p_2 q + d^{byz} \sin \tilde p_2 q) \times
 \delta ^{zx} \sin \tilde p_3(q+p_1)
\eea
which can be simplified (by using \eqref{I1}, \eqref{I2} and
\eqref{sin}) and rewritten as
\beq
V_3 = - g^3 \sqrt{2\over N} N (2- {4\over N^2}) {1\over 4} \sin \tilde
p_3 p_1 \delta ^{ab} +\ \mbox{other terms} \label{V3}
\eeq
In addition there are two other diagrams which correct the
$SU(N)-SU(N)-U(1)$
vertex. In one of the diagrams there are two $U(1)$ bosons and one $SU(N)$
bosons which flow in the triangle (fig. 3a) and in the other there
are two $SU(N)$ bosons and one $U(1)$. The two diagrams contributes
the same. Their contribution is
\bea
 V'_3 & = & 2\times g^3 {({2\over N})}^{3\over 2}
 \sin \tilde p_1 q \sin \tilde p_2 q \sin \tilde p_3(q+p_1) \delta
^{ab} \nonumber  \\
 & = & -2\times g^3 \sqrt{2\over N} {2\over N} {1\over 4}\sin \tilde p_3 p_1 \delta ^{ab}
+\ \mbox{other terms}
 \label{V33}
\eea
The contribution $V'_3$ exactly compensate the ${1\over N^2}$ part in
 \eqref{V3}. Hence the needed counter term is
\beq
\delta _1 ^{(N-N-1)} = {g^2 N \over (4\pi)^2} 
\times {2\over 3} \times {2\over \epsilon} ,  \label{CT1NN}
\eeq
exactly as \eqref{CT111} and \eqref{CTNNN}.

The 'other terms' in eqs.\eqref{V3},\eqref{V33} leads to finite terms
which take exactly the same form as \eqref{finiteV}.

The calculation of the 4-gluon vertices is straightforward, though
tedious. Adding matter in the adjoint representation is also
straightforward. The counter terms which are needed in all these cases
 are exactly the same as the ones which are needed in ordinary $SU(N)$ theory. 

\section{Renormalizability and gauge invariance}

In the previous sections we calculated the counter terms which are
needed to renormalize the theory. Since we are dealing with a gauge 
theory, gauge symmetry imposes some constraints on the various counter
terms. In ordinary Yang-Mills theory the three gluons vertex and the
four gluons vertex are multiplied by $g$ and $g^2$ respectively. Gauge
invariance tells us that the two couplings should be the same - also 
 at the quantum level. In the present case the situation is even more 
involved. A-priori, there are two types of propagators with different
wave functions renormalization. There are also three types of
vertices (even four, if we consider the $f^{abc} \cos$ and the
$d^{abc} \sin$ parts
of the $SU(N)$ vertex as two
 independent vertices). 

Gauge invariance imposes the following relations, at one loop
\beq
\delta _1 ^{(1-1-1)} - {3\over 2} \delta _3 ^{(1-1)} =
\delta _1 ^{(N-N-N)} - {3\over 2} \delta _3 ^{(N-N)} =  
\delta _1 ^{(N-N-1)} - \delta _3 ^{(N-N)} - {1\over 2} \delta _3
^{(1-1)}.
\label{GI}
\eeq
We have found that in fact all $\delta _1 ^i$ are equal and $\delta _3
^i$ are equal. Clearly, \eqref{GI} is satisfied. 

The calculation of the beta function is also straightforward. The ${2\over
\epsilon}$ in the expressions for $\delta$ should be replaced by $\log
{\Lambda^2\over \mu^2} $ and the beta function is computed by 
\beq
 \beta (g) = g \mu {\partial \over \partial \mu} (-\delta _1 + {3\over
2} \delta _3).
\eeq
The result is
\beq 
 \beta (g) = -{g^3 \over (4\pi)^2} {11\over 3} N,
\eeq 
as expected\cite{MRS}. Note that our result for the $U(1)$ case differs by a
factor of $2$ from \cite{MaSa} due to a different definition of the
$U(1)$ coupling.  

Let us comment about various limits and some special cases.
In contrast to the commutative theory, where the $U(N)$ theory
 contains two couplings: a $U(1)$ coupling which doesn't run and an
 asymptotically free $SU(N)$ coupling, we showed that non-commutative 
 theory can (and as we shall see in a moment - must) contain a single
 coupling. The theory is asymptotically free and the value of the beta
function is independent of $\theta$. It is a bit unusual at first
sight, since the commutative
 theory should be a limit of the non-commutative theory. However, this limit
is singular. As long as $\theta$ is non-zero the $U(1)$ coupling runs,
 independently of the value of $\theta$ and exactly as the 
$SU(N)$ coupling. When $\theta$ is zero, the $U(1)$ is frozen. Thus
though the Feynman rules of the non-commutative theory are smooth in
$\theta$, the renormalization procedure makes the limit singular. 

The planar limit is also interesting. It was suggested
\cite{Filk,BS} that the planar limit of the non-commutative theory is
the ordinary theory. However, the planar limit of the $U(1)$ theory is
not the ordinary commutative theory but rather an interacting theory
and the counter term \eqref{CT11} is still needed.
Similarly, the planar limit of the $U(N)$ theory is not the
planar limit of the $SU(N)\times U(1)$ ordinary theory.
In order to be more precise, let us give an example which clarifies
 the difference between the commutative and the non-commutative
planar theories. Correlation functions which involve {\em only}
${\rm tr}\ F_{\mu\nu}$ would yield trivial answers in the commutative theory,
since the $U(1)$ part is decoupled and free. On the
 other hand, such correlation functions are highly non-trivial in the planar 
non-commutative case.

Another remark is about the $SU(N)$ theory. It was argued in
\cite{Matsubara} (see also \cite{MSSW,Terashima} and \cite{AD} for a
derivation from string theory) that the non-commutative
 version of this theory is not consistent, 
since the closure of the Moyal commutator is violated.
Here we find another evidence for the inconsistency of the $SU(N)$
theory. Had we ignored the $U(1)$ part of the theory, we would have
found that the value of the beta function is gauge dependent. In order
to see that one should calculate the counter terms in a general gauge
and to observe that a ${1\over N^2}$ gauge dependent piece is left
in the beta function of the $SU(N)$ theory.

Finally let us comment about the ${\cal N}=4 $ theory. Since this
theory is finite in the ordinary commutative case, it seems that for
this specific case the limit $\theta \rightarrow 0$ is smooth. Let us
focus on the planar theory first. Both the
$U(1)$ and the $SU(N)$ gauge couplings take their classical value and
no counter terms are needed. Therefore the $\theta \rightarrow 0$ limit
is the same as the $\theta = 0$ theory. The non-planar sector of the
theory is more subtle. The finite UV effects are manifestly singular
in $\theta$. It was suggested in \cite{MST}, that in the specific case
 of ${\cal N}=4 $ these contributions cancel and thus also this sector
of the theory is smooth in $\theta$.

 We would like to note that there
is another class of theories which are UV finite
 \cite{Armoni} and maybe even smooth
in $\theta$. These are the orbifold truncations of ${\cal N}=4$. These
 theories share the same planar diagrams as ${\cal N}=4$ \cite{BJ}.
Therefore, this sector of the theory is finite. 
The non-planar sector is anyways UV
finite in non-commutative theories. 
Moreover, since these theories admit Bose-Fermi degeneracy, it is
likely that the non-planar contributions cancel as in the ${\cal N}=4$
case.

\newpage

\Acknowledgements

I would like to thank C. Angelantonj, I. Antoniadis, E. Gardi, 
F. Hassan and R. Minasian for discussions and comments. 
This research was supported in part by EEC under TMR contract 
ERBFMRX-CT96-0090.

\newpage

\section{Appendix A - Feynman rules for the non-commutative $U(N)$ Yang-Mills theory}
The non-commutative Yang-Mills action including gauge fixing
and ghosts takes the following form
\beq
S= \int d^4 x\  {\rm tr} \ \left (-{1\over 2} F^{\mu \nu} \star F  _{\mu \nu}
+
 \xi (\partial ^\mu A _\mu)^2 - 
 \bar c \star \partial ^\mu D_\mu c 
+ \partial ^\mu D_\mu c \star \bar c \right )
\eeq  
We use the Feynman-'t Hooft gauge $\xi=1$. 
\begin{figure}[H]
  \begin{center}
\mbox{\kern-0.5cm
\epsfig{file=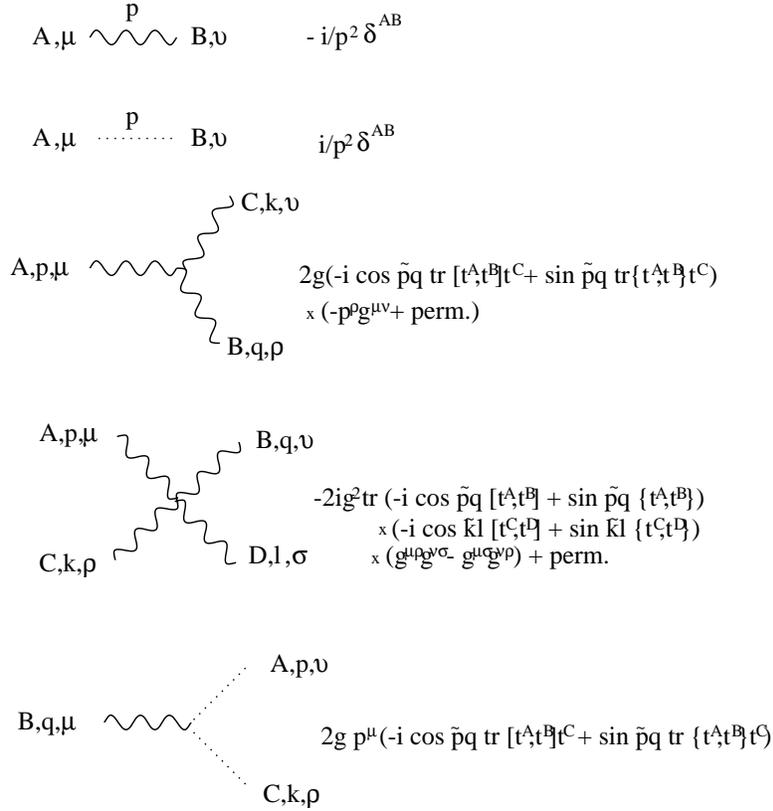,width=10.0true cm,angle=0}}
\label{rules1}
  \end{center}
\caption{Feynman rules. 
Wavy lines and dotted lines denote gluons and ghosts, respectively.
Capital letters and small letters denote $U(N)$ indices and momenta.}

\end{figure}

\newpage

\section{Appendix B - $SU(N)$ identities}
In this section we derive $SU(N)$ identities which are used in the
paper. 

{\bf Identity A} $f^{axy} f^{bxy} = N \delta ^{ab}$.

We denote the adjoint representation by capital letters. We use 
${\rm tr} \ T^a T^b = N \delta ^{ab}$. Also $T^a _{xy} = -if^{axy}$.
Therefore $T^a _{xy} T ^b _{yx} = -i f^{axy}\times -i f^{byx} = f^{axy}
f^{bxy} = N \delta ^{ab}$.

{\bf Identity B} $d^{axy} d^{bxy} = (N -{4\over N}) \delta ^{ab}$.

\beq
 {\rm tr} \ t^x t^x t^a t^b = {N^2-1\over 2N} {1\over 2} \delta ^{ab}
 \label{A1} .
\eeq
By using $ t^a t^b = {1\over 2} (i f ^{abc} t^c + {1\over N} \delta
 ^{ab} + d^{abc} t^c )$, \eqref{A1} reads
\bea
 & & = {\rm tr} \ {1\over 2} (if^{xay} t^y + {1\over N} \delta ^{xa} + d^{xay} t ^y) t^b
 t^x \nonumber \\
 & & = {1\over 2} (if^{xay} +d^{xay}) {\rm tr} \ t^y t^b t^x + {1\over 4}
 {1\over N} \delta ^{xa} \delta ^{bx} \nonumber \\
 & & = {1\over 8} (f^{axy} f^{bxy} + d^{axy} d^{bxy} ) + {1\over 4N}
 \delta ^{ab} \label{A2}
\eea
Thus \eqref{A1} and \eqref{A2} leads to 
\beq
f^{axy} f^{bxy} + d^{axy} d^{bxy} = (2N - {4\over N}) \delta ^{ab},
\eeq
and by using identity A, identity B is proven.   

{\bf Identity C} 
\bea
 & & f^{axy} f^{byz} f^{czx} - f^{axy} d^{byz} d^{czx} \nonumber \\
 & & -  d^{axy} f^{byz} d^{czx}  -  d^{axy} d^{byz} f^{czx} =
2N(1-{3\over N^2}) f^{abc} \\
& & d^{axy} d^{byz} d^{czx}  - d^{axy} f^{byz} f^{czx} \nonumber \\
& & -f^{axy} d^{byz} f^{czx} - f^{axy} f^{byz} d^{czx} =
2N(1-{3\over N^2}) d^{abc} 
\eea
we begin with 
\beq
 {\rm tr} \ t^x t^x t^a t^b t^c = {N^2 -1 \over 2N} {1\over 4} (if^{abc} +
 d^{abc}). \label{A3}
\eeq
 Eq. \eqref{A3} can be written also as follows
\bea
 & & = {\rm tr} \ {1\over 2} (if^{xay} t^y + {1\over N} \delta ^{xa} + d^{xay}
 t^y) t^b t^c t^x  \nonumber \\
 & & = {1\over 2}(if^{xay} +d^{xay} ) {\rm tr} \ t^y t^b t^c t^x + {1\over 2N}
{\rm tr} \ t^b t^c t^a  \nonumber \\
 & & = {1\over 16} (if^{xay} + d^{xay} )(if^{ybz} +
 d^{ybz})(if^{zcx}+d^{zcx})+{2\over 8N} (if^{abc}+d^{abc}). \label{A4}
\eea
By equating \eqref{A3} and \eqref{A4} we arrive at
\beq
(-if^{axy} + d^{axy})(-if^{byz} +d^{byz})(-if^{czx}
 +d^{czx})=(2N-{6\over N})(if^{abc} +d^{abc}). \label{A5}
\eeq
The real and imaginary parts of \eqref{A5} prove identity C.

\end{document}